\documentclass[
    amsmath,
    amssymb,
    aps,
    prl,
    reprint, 
]{revtex4-2}

\usepackage[utf8]{inputenc}
\usepackage[english]{babel}

\usepackage{
    amsfonts,
    booktabs,
    csquotes,
    graphicx,
    subcaption,
    url,
}
\usepackage[pdfusetitle]{hyperref} 
\hypersetup{colorlinks=false,pdfborder={0 0 0}} 

\usepackage{
    pgfplots,
    tikz,
}
\usetikzlibrary{
    arrows.meta,
    bending,
    calc,
    positioning,
}
\usepgfplotslibrary{groupplots}

\pgfplotsset{
    compat=1.16, 
    xlabel near ticks, 
    ylabel near ticks,
    small, 
}

\newcommand{\di}[1]{\textrm{d}#1}

\newcommand{\dz}{\di{z}}

\newcommand{\deriv}[2]{\frac{\textrm{d}#1}{\textrm{d}#2}}

\newcommand{\ddx}[1]{\deriv{#1}{x}}

\newcommand{\ddxInline}[1]{\di{#1}/\di{x}}

\newcommand{\parens}[1]{\left ( #1 \right )}

\newcommand{\bigParens}[1]{\big ( #1 \big )}

\newcommand{\boltz}{k_\textrm{B}}
\newcommand{\muBulk}{\mu_b}
\newcommand{\muEff}{\mu^*}
\newcommand{\muInt}{\mu_I}

\newcommand{\software}[1]{\MakeLowercase{\textsc{#1}}}

\begin{document}

\begin{abstract}
\noindent
Interfacial tension gradients drive flow along liquid--liquid interfaces in a process known as the Marangoni effect. Such gradients can be caused by surfactants, which has been extensively studied in the literature. Less is known of its nanoscale properties, where molecular interfaces display properties separate from bulk liquid matter such as interfacial viscosity. In this report we study the molecular solutal Marangoni effect using molecular dynamics simulations. We show that molecular interfacial effects are important and should be accounted for in nanofluidic regimes. Hydrodynamic models can be extended with effective terms which include them. 
\end{abstract}

\title{How molecular effects affect solutal Marangoni convection}

\author{Petter Johansson}
\email{pjohansson@univ-pau.fr}
\author{Guillaume Galliéro}
\affiliation{Laboratoire des Fluides Complexes et Leurs Réservoirs, Université de Pau et des Pays de l’Adour, E2S UPPA, CNRS, Total, LFCR, Pau, France}

\author{Dominique Legendre}
\affiliation{Institut de Mécanique des Fluides de Toulouse (IMFT), Université de Toulouse, CNRS-INPT-UPS, Toulouse, France}

\date{\today}
\maketitle

\noindent
The Marangoni effect (also called the Gibbs--Marangoni effect) describes liquid transport along liquid--liquid interfaces from regions of low-to-high surface or interfacial tension \cite{Scriven1960,deGennes2004}. It is a powerful effect, requiring only small interfacial tension gradients to create strong convective flows. Such gradients arise naturally by the addition of a surfactant at a liquid interface, as has been studied in various configurations using both experiments \cite{Bazazi2020} and continuum fluid dynamics simulations \cite{Atasi2018}. This solutal Marangoni effect plays a role in many phenomena, in particular foam and emulsion formation and evolution \cite{Walstra1993}. 

The phenomena is well understood from the macroscopic point of view, where the induced flow velocity at the interface is linear to the interfacial tension gradient \cite{Scriven1960}. However, less is known on possible microscopic effects on such phenomena when the width of the interface is not negligible compared to the size of the bubble or droplet, such as when considering nanoemulsion, nanobubbles or nanodroplets \cite{Firouzi2017,Espitia2019,Hack2021}. In this report we will show how molecular details influence the solutal Marangoni effect of a model liquid--liquid interface populated with surfactant molecules.

Molecular dynamics simulations have shown to be an efficient complementary tool to experiments to better understand flows at the nanoscale, in particular close to interfaces \cite{Eijkel2005,Bocquet2010}. Regarding solutal Marangoni flow, molecular dynamics simulations have been used to confirm that nanoscopic interfacial tension gradients create molecular flows \cite{Imai2017,Imai2019} and that the right force leading to Marangoni flow is related to chemical potential gradients \cite{Liu2017}. 

Less is known of how surfactants and molecular details affect the creation of a flow at the nanoscale. Interfaces themselves are of non-zero width and display properties separate from the bulk phases, especially in the presence of surfactants.
In particular it has been shown that the viscous dissipation across a liquid--liquid interface\,---\,with and without surfactant molecules\,---\,can be characterized by an interfacial viscosity (i.e.~a partial slip at the liquid--liquid interface) \cite{Galliero2010,Poesio2017,Zhan2020}
\begin{equation}
    \label{eq:interfacial-viscosity}
    \muInt = \frac{\tau_{xz} w}{\Delta u_{x,I}}
\end{equation}
where $\tau_{xz}$ is a tangential shear over the interface (which is normal to $z$), $w$ the interface width and $\Delta u_{x,I}$ the change in velocity over that width. While this contribution to the dissipation in a system is negligible on a macroscopic scale it is important to account for as we approach the nanoscale. 

With this work we use molecular dynamics simulations to investigate the Marangoni effect on a molecular scale. In particular the varying interfacial viscosity is shown to have a large influence as molecular length scales are approached. Such a contribution can be modeled using an effective measure, which does not require an explicit model of the interface.


\begin{figure}
    \centering
    \begin{subfigure}[b]{0.4\columnwidth}
        \centering
        \includegraphics{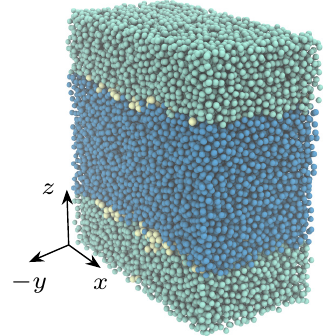}
        \subcaption{}
    \end{subfigure}
    \begin{subfigure}[b]{0.58\columnwidth}
        \includegraphics{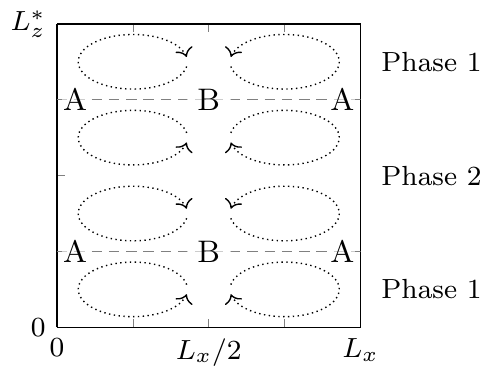}
        \subcaption{\label{fig:system-definition-sketch}}
    \end{subfigure}
    \newline
    \begin{subfigure}{\columnwidth}
        \centering
        \includegraphics{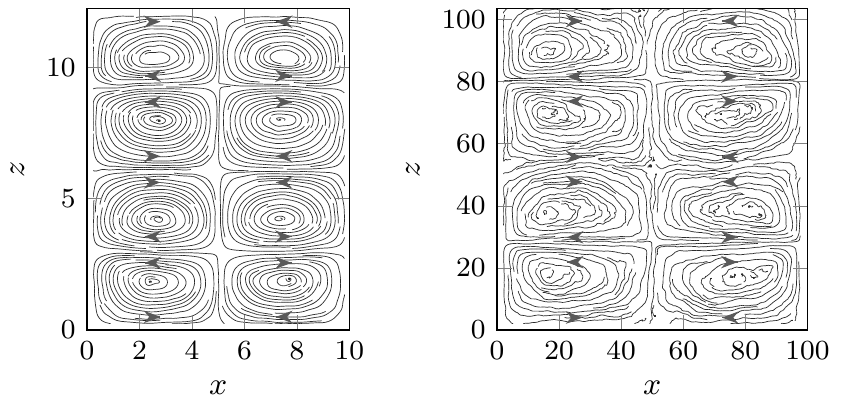}
        \subcaption{\label{fig:system-definition-streamlines}}
    \end{subfigure}
    \caption{\label{fig:system-definition} a)~Simulated two-phase system with surfactants at interfaces (white). b)~2D view with low (A) and high (B) surfactant concentration zones marked. Resulting flow vortices are sketched. c)~Streamline plots of developed flow in systems of input size $10\times10$ and $100\times100$. Figures are created with \software{VMD}, \software{Matplotlib} and \software{PGFPlots} \cite{Humphrey1996,Hunter2007,Feuersanger2021}.}
\end{figure}

To start we consider a two-phase system with liquid phases $1$ and $2$ shown in figure \ref{fig:system-definition}. The phases are immiscible and separated along the $z$ axis by two interfaces. Both phases consist of dimer Lennard-Jones molecules with identical atoms of mass $m$. Intermolecular interactions between atoms in phase $i$ and $j$ are given by the Lennard-Jones potential
\begin{equation}
    \label{eq:lennard-jones-potential}
    U_{ij}(r) = 4 \varepsilon_{ij} \parens{
        \frac{\sigma_{ij}^{12}}{r^{12}} - \frac{\sigma_{ij}^6}{r^6}
    }
\end{equation}
where $r$ is the distance between the atoms and $\varepsilon_{ij},\sigma_{ij}$ the interaction parameters. Phases $1$ and $2$ are made identical by setting $\varepsilon_{11} = \varepsilon_{22} \equiv \varepsilon$ and $\sigma_{11} = \sigma_{22} \equiv \sigma$ and immiscible by setting the cross-interaction strength $\varepsilon_{12} = 0.5\varepsilon$. Internal harmonic bonds with strength $k = 1000\varepsilon$ and distance $\sigma$ keep the molecules together. Surfactant molecules are created as identical dimers where one atom is of species $1$ and the other of species $2$: the first is attracted to phase 1, the other to phase 2. 

Simulations are performed and results presented in Lennard-Jones reduced units \cite{AllenTildesley2017} with $\varepsilon = \sigma = m = \boltz = 1$, where $\boltz$ is the Boltzmann constant. The temperature is $T = 0.8$ and bulk atom number density in the liquid phases is $\rho_b = 0.87$. Without surfactants the interfacial tension of the two-phase interface is $\gamma_0 = 1.55$ (see below). Using non-equilibrium molecular dynamics \cite{Bordat2002,Galliero2005,Galliero2009} the shear viscosity of the bulk is measured to be $\muBulk = 4.4$. Simulations are performed with \software{Gromacs 2020} \cite{Abraham2015,Pall2020} using a leap frog integrator with a time step $dt = 0.002$ and interaction cut-off range $r_c = 3.5$. Periodic boundary conditions are enforced along all dimensions. Temperature is controlled with a velocity rescaling thermostat with coupling time $t_c = 5.0$ \cite{Bussi2007}.

To characterize the influence of surfactant concentration $\rho_I$ at the interface we measure the interfacial tension $\gamma(\rho_I)$, interface width $w(\rho_I)$ and interfacial viscosity $\muInt(\rho_I)$. These measurements are done in three steps. 

First, surfactant molecules are inserted at the two interfaces and an equilibration simulation is run to achieve a consistent bulk liquid density $\rho_b$. The box size is $L_x = L_y = 20$ along $x$ and $y$, and $L_z = 40$ along $z$ before the surfactant is added. The box size including the interfaces is $L_z^* = L_z + 2w$, where $w$ is up to a few $\sigma$. 

Second, an NVT simulation of $10^7$ steps is run. From this simulation, the surfactant concentration (atom number density) $\rho_I$ is defined by matching a Gaussian distribution $\rho(z) = \rho_I e^{-a(z - z_i)^2}$ to each interface at position $z_i$ and taking the mean. The interfacial tension $\gamma$ is simultaneously calculated from the stress tensor fluctuations of the diagonal terms across the simulation box, dividing by the number of interfaces and averaging over the entire simulation time $t$ \cite{AllenTildesley2017}:
\begin{equation}
    \label{eq:interfacial-tension}
    \gamma = \frac{1}{2} \left \langle \int_0^{L_z^*} \dz\, \bigParens{
        P_{zz}(z,t) - P_{tt}(z, t)
    } \right \rangle_t 
\end{equation}
where $P_{ij}$ is the pressure tensor and $P_{tt}(z, t) = \frac{1}{2} (P_{xx}(z, t) + P_{yy}(z, t))$.

\begin{figure}
    \centering
    \includegraphics{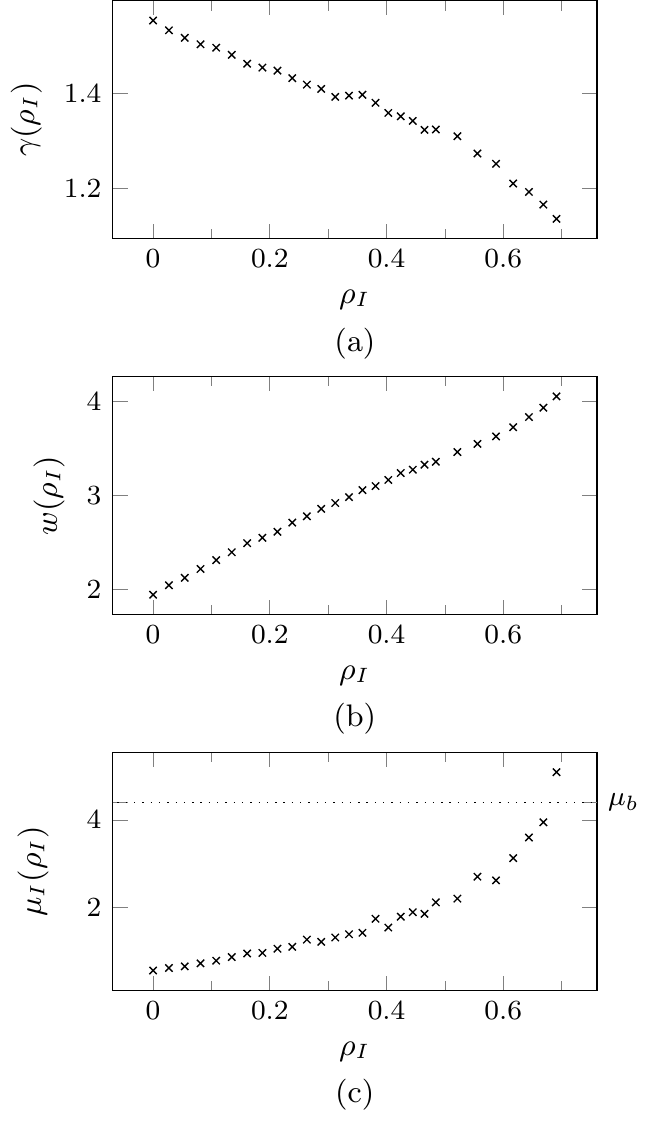}
    \caption{\label{fig:interface-properties-per-surfactant-conc} Interfacial tension $\gamma$, width $w$ and viscosity of the interface $\muInt$ for various surfactant concentrations $\rho_I$ at the liquid--liquid interface.}
\end{figure}

Finally, following \cite{Galliero2010} a shear $\tau_{xz}$ is created in the system using the same method as when measuring $\muBulk$. The shear simulation is run for $2 \cdot 10^6$ steps to create a steady state, after which data is collected over $2 \cdot 10^7$ steps. We then calculate the interfacial width $w$ as the distance between where the density profiles of phase 1 and 2 reach $\rho = 0.95\rho_b$ which allows us to measure $\Delta u_{x,I}$ and calculate $\muInt$ using \eqref{eq:interfacial-viscosity}. See figures S1 and S2 in the supplementary materials \cite{SupplementaryMaterial} for visual definitions of $w(\rho_I)$ and $\Delta u_{x,I}(\rho_I)$.

Repeating these measurements for a range of $\rho_I$ we can characterize how the surfactant density influences the interfacial properties. This is reported in figure \ref{fig:interface-properties-per-surfactant-conc} for $\rho_I \in [0, 0.7]$. An interesting observation is that while the interface is only a few atoms wide when no surfactants are added, the interfacial viscosity drops by an order of magnitude compared to the bulk. Note also that as the interface is saturated with surfactants a separate bulk phase develops with viscosity similar to the bulk phases. The concentrations considered in the next section are below this point, with interface concentrations $\rho_I < 0.5$.

The Marangoni effect describes the flow created by a varying interfacial tension $\gamma$ over an interface. To induce an interfacial tension gradient along $x$ at our interfaces we employ a simple non-equilibrium scheme. For each interface we define an edge zone $A$ centered at $x = 0$ and a center zone $B$ at $x = L_x / 2$ (figure \ref{fig:system-definition-sketch}). Each zone is of size $l_x \times l_z$ along $x$ and $z$ and spans the entire width in~$y$. 

At every $N$ step of the simulation we check whether a surfactant molecule exists in zone $A$ and if so exchange its flavor with a liquid molecule in zone $B$. This scheme mimics that of \citeauthor{Liu2017} \cite{Liu2017} except focused at the interfaces and applied to dimers instead of monomers. The position and momentum of the molecule remaining in each zone is not changed, only the molecule type. Since all molecules are of equal size and mass we preserve local momentum. If the exchange frequency $N$ is sufficiently high a gradient $\ddxInline{\rho_I}$ forms along the interface (see figure \ref{fig:properties-along-interface}a and figure S3 in the supplementary material \cite{SupplementaryMaterial}).

Simulation systems are prepared and equilibrated following the steps of the previous section for varying sizes $L_x = L_z \in [10, 100]$. The size along $y$ is kept at a constant $L_y = 20$. The exchange step frequency is $N = 500$ and the exchange zone sizes are $l_x = l_z = 2$. For each system size we start 4 independent simulations, each of which runs for between $5 \cdot 10^7$ and $5 \cdot 10^8$ steps. We discard data from the first $2 \cdot 10^6$ steps to allow for the flow field in the system to develop. This is verified by comparing to data from the second half of the simulation. Finally, we collect the average flow field (mass and velocity) from the simulation in bins of size $0.25 \times 0.25$ along $x$ and $z$.


\begin{figure}
    \centering
    \includegraphics{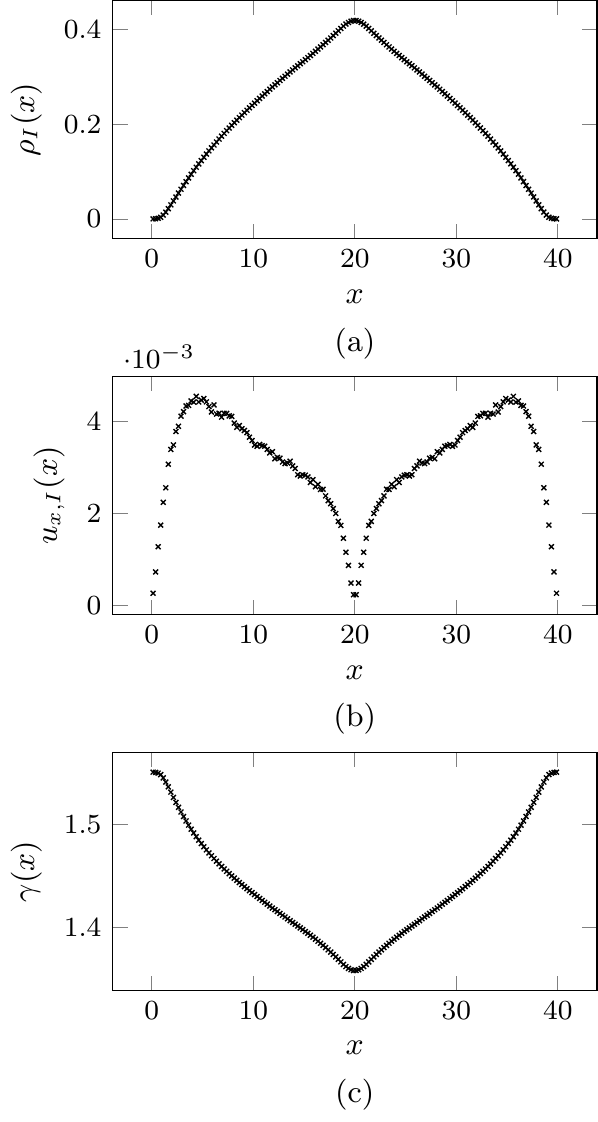}
    \caption{\label{fig:properties-along-interface} Surfactant concentration $\rho_I$, flow velocity $u_{x,I}$ and interfacial tension $\gamma$ along the interfaces for a system of size $40 \times 40$.}
\end{figure}

After starting the simulations a flow rapidly develops throughout the system, forming flow vortices exemplified in figure \ref{fig:system-definition-streamlines}. We see that the flow vortices are not centered in each cell center. They trend towards a point that is two-thirds along the interface with a finite shift towards the center that is noticeable only for the smallest systems. We choose this vortex center point as the reference axis along $z$ for comparing field data, since the flow will be most similar along it for all systems. See figures S4 and S5 in the supplementary material \cite{SupplementaryMaterial} for more streamlines and a view of this reference axis along the center points.

Using the binned flow field data we measure the surfactant density $\rho_I(x)$ and velocity $u_{x,I}(x)$ along the interface. These are shown along with $\gamma(x)$ in figure \ref{fig:properties-along-interface} for a system of size $L_x = L_z = 40$, where the data has been symmetrized around the box center $x = L_x / 2$ and $u_{x,I}$ is positive for a flow pointing away from the center. It is worth noting that $w(x)$ and $\muInt(x)$ vary linearly with $\rho_I(x)$ for these densities and result in similar profiles (see figure S6 of the supplementary material \cite{SupplementaryMaterial}). For the velocity we see a linear change in the center of the interface, with large changes at the edges where the velocity shifts from being transverse to normal to the interface.

Are these results well described by hydrodynamic modeling? Not if we neglect to model the interface. Marangoni convection velocity is related to the interfacial tension gradient $\nabla \gamma$ and viscosity $\mu$: $u_x \propto \nabla \gamma / \muBulk$, if we take the bulk shear viscosity $\muBulk$. But in figure \ref{fig:properties-along-interface}\hyperlink{fig:properties-along-interface-velocity}{b}--\hyperlink{fig:properties-along-interface-surften}{c} we see a velocity gradient at the center, where the interfacial tension gradient is constant. The effect remains even as we double the system size along $x$ only (figure S6 in the supplementary material \cite{SupplementaryMaterial}), thus it is not due to the finite size of the system or to hydrodynamic effects.

\begin{figure}
    \centering
    \begin{subfigure}[b]{0.48\columnwidth}
        \includegraphics{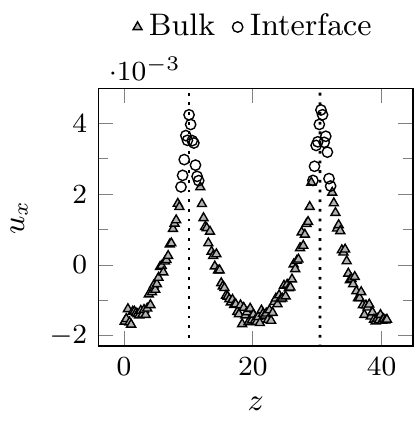}
        \subcaption{\label{fig:velocity-profiles-full}}
    \end{subfigure}
    ~
    \begin{subfigure}[b]{0.48\columnwidth}
        \includegraphics{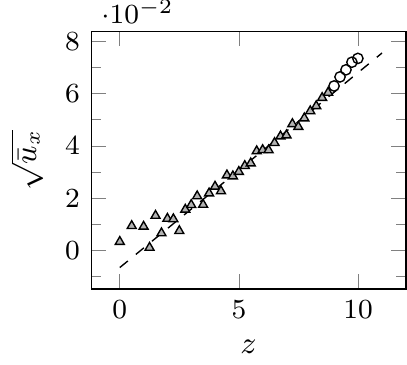}
        \subcaption{\label{fig:velocity-profiles-sqrt}}
    \end{subfigure}
    \caption{\label{fig:velocity-profiles} a)~Velocity profile $u_x(z)$ for system size $40 \times 40$. Interface centers are marked. b)~Square-root plot where $\bar{u}_x = u_x(z) - u_0$ adjusts the profile to its approximate minima. Linear fit of center bulk values drawn as a dashed line.}
\end{figure}

To model this velocity change we have to account for the interface itself. So what happens inside of it? In figure \ref{fig:velocity-profiles-full} we extract the velocity profile $u_x(z)$ through the reference axis of a $40 \times 40$ size system and denote the bulk and interface with different markers. We see that $u_x(z)$ is quadratic in each bulk phase, with a minima at the center and maximum at the interface. This is apparent by adjusting the minima to 0 and taking the square root, which results in a linear profile through the center bulk (figure \ref{fig:velocity-profiles-sqrt}). However, at the interface the linear slope changes, which indicates a change in viscosity inside the interface. This confirms the influence of interfacial viscosity $\muInt$ \eqref{eq:interfacial-viscosity}. See figures S7 and S8 in the supplementary material for profiles of more system sizes \cite{SupplementaryMaterial}.

We now consider how to describe the flow using hydrodynamics. In the supplementary material \cite{SupplementaryMaterial} we derive a hydrodynamic description for the interface velocity $u_{x,I}$ (eq.~S27):
\begin{equation}
    \label{eq:marangoni-flow-velocity}
    u_{x,I}(x) = \frac{\alpha H}{\mu} \ddx{\gamma} \, ,
\end{equation}
where $H$ is the height of the flow-reverse-flow vortex along $z$, $\mu$ is the shear viscosity and $\alpha$ a coefficient which is $\alpha = 1/6$ for simplified flow configurations similar to those we observe at the vortex center axis in all our systems. If the shear dissipation is correctly modeled, equation~\ref{eq:marangoni-flow-velocity} should yield a constant $\alpha$ for our measured $u_{x,I}$, $\ddxInline{\gamma}$ and $\mu$ at these points. 

\begin{figure}
    \centering
    \includegraphics{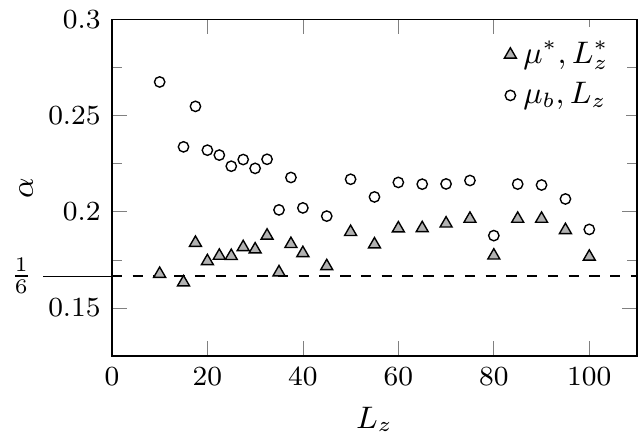}
    \caption{\label{fig:alpha-coefficient} $\alpha$-coefficients for a range of system sizes $L_x = L_z$.}
\end{figure}

With this in mind we compute $\alpha$ in two ways for our full range of system sizes (see figure \ref{fig:alpha-coefficient}): First, using only the bulk phase dissipation, by setting $\mu = \muBulk$ and $H = L_z / 4$ since there are four vortices along $z$. Here $\alpha$ changes dramatically for systems with $L_z < 40$, where the interface is prominent. For larger systems $\alpha$ is around 30\% higher than $1/6$. 

Second, we include the interfacial dissipation by calculating the effective viscosity $\muEff(x)$ using a harmonic average
\begin{equation}
    \label{eq:effective-viscosity}
    \frac{w(x) + L_b(x)}{\muEff(x)} = \frac{w(x)}{\muInt(x)} + \frac{L_b(x)}{\muBulk}
\end{equation}
where $L_b(x) = L_z^* / 2 - w(x)$ is the width of the bulk phases. Setting $H = L_z^* / 4$ and $\mu = \muEff$ to estimate $\alpha$ for our systems we obtain an improved agreement with the description. $\alpha$ is now constant and close to $1/6$.

We make two conclusions: First, the finite interface must be accounted for to model mesoscopic systems. This is shown by the inability to model the velocity gradient $u_{x,I}$ without accounting for the varying interface properties. Second, the interface viscosity $\muInt$ \eqref{eq:interfacial-viscosity} which is included in the effective viscosity $\muEff$ \eqref{eq:effective-viscosity} is a good measure of the dissipation inside an interface. While this report has focused on solutal Marangoni convection, this has implications for any systems which model nanoscale flows with interfaces.

A few complications are of note. The assumptions made to derive \eqref{eq:marangoni-flow-velocity} are simplified. In particular, our streamlines are not perfectly parallel to the interface along the vortex axis, but slightly tilted (figure \ref{fig:system-definition-streamlines}). This leads to estimated coefficients which are slightly higher than $\alpha = 1/6$. The tilt decreases for our smaller systems, where the measured $\alpha$ coefficients are closer to $1/6$. This supports the description as being qualitatively correct. Furthermore, our molecular modeling is very simple. Further study into interfacial properties using more realistic liquid models are required to understand their real implications for nanoscale flows.

\begin{acknowledgments}
We gratefully acknowledge Institut Carnot ISIFOR for the post-doctoral grant awarded to one of us (PJ). We also thank UPPA for providing computational facilities at the Pyrene cluster.
\end{acknowledgments}


\begin{thebibliography}{27}%
\makeatletter
\providecommand \@ifxundefined [1]{%
 \@ifx{#1\undefined}
}%
\providecommand \@ifnum [1]{%
 \ifnum #1\expandafter \@firstoftwo
 \else \expandafter \@secondoftwo
 \fi
}%
\providecommand \@ifx [1]{%
 \ifx #1\expandafter \@firstoftwo
 \else \expandafter \@secondoftwo
 \fi
}%
\providecommand \natexlab [1]{#1}%
\providecommand \enquote  [1]{``#1''}%
\providecommand \bibnamefont  [1]{#1}%
\providecommand \bibfnamefont [1]{#1}%
\providecommand \citenamefont [1]{#1}%
\providecommand \href@noop [0]{\@secondoftwo}%
\providecommand \href [0]{\begingroup \@sanitize@url \@href}%
\providecommand \@href[1]{\@@startlink{#1}\@@href}%
\providecommand \@@href[1]{\endgroup#1\@@endlink}%
\providecommand \@sanitize@url [0]{\catcode `\\12\catcode `\$12\catcode
  `\&12\catcode `\#12\catcode `\^12\catcode `\_12\catcode `\%12\relax}%
\providecommand \@@startlink[1]{}%
\providecommand \@@endlink[0]{}%
\providecommand \url  [0]{\begingroup\@sanitize@url \@url }%
\providecommand \@url [1]{\endgroup\@href {#1}{\urlprefix }}%
\providecommand \urlprefix  [0]{URL }%
\providecommand \Eprint [0]{\href }%
\providecommand \doibase [0]{https://doi.org/}%
\providecommand \selectlanguage [0]{\@gobble}%
\providecommand \bibinfo  [0]{\@secondoftwo}%
\providecommand \bibfield  [0]{\@secondoftwo}%
\providecommand \translation [1]{[#1]}%
\providecommand \BibitemOpen [0]{}%
\providecommand \bibitemStop [0]{}%
\providecommand \bibitemNoStop [0]{.\EOS\space}%
\providecommand \EOS [0]{\spacefactor3000\relax}%
\providecommand \BibitemShut  [1]{\csname bibitem#1\endcsname}%
\let\auto@bib@innerbib\@empty
\bibitem [{\citenamefont {Scriven}\ and\ \citenamefont
  {Sternling}(1960)}]{Scriven1960}%
  \BibitemOpen
  \bibfield  {author} {\bibinfo {author} {\bibfnamefont {L.~E.}\ \bibnamefont
  {Scriven}}\ and\ \bibinfo {author} {\bibfnamefont {C.~V.}\ \bibnamefont
  {Sternling}},\ }\href {https://doi.org/10.1038/187186a0} {\bibfield
  {journal} {\bibinfo  {journal} {Nature}\ }\textbf {\bibinfo {volume} {187}},\
  \bibinfo {pages} {186} (\bibinfo {year} {1960})}\BibitemShut {NoStop}%
\bibitem [{\citenamefont {de~Gennes}\ \emph {et~al.}(2004)\citenamefont
  {de~Gennes}, \citenamefont {Brochard-Wyart},\ and\ \citenamefont
  {Qu{\'{e}}r{\'{e}}}}]{deGennes2004}%
  \BibitemOpen
  \bibfield  {author} {\bibinfo {author} {\bibfnamefont {P.-G.}\ \bibnamefont
  {de~Gennes}}, \bibinfo {author} {\bibfnamefont {F.}~\bibnamefont
  {Brochard-Wyart}},\ and\ \bibinfo {author} {\bibfnamefont {D.}~\bibnamefont
  {Qu{\'{e}}r{\'{e}}}},\ }\href {https://doi.org/10.1007/978-0-387-21656-0}
  {\emph {\bibinfo {title} {{Capillarity and Wetting Phenomena}}}}\ (\bibinfo
  {publisher} {Springer New York},\ \bibinfo {address} {New York, NY},\
  \bibinfo {year} {2004})\BibitemShut {NoStop}%
\bibitem [{\citenamefont {Bazazi}\ and\ \citenamefont
  {Hejazi}(2020)}]{Bazazi2020}%
  \BibitemOpen
  \bibfield  {author} {\bibinfo {author} {\bibfnamefont {P.}~\bibnamefont
  {Bazazi}}\ and\ \bibinfo {author} {\bibfnamefont {S.~H.}\ \bibnamefont
  {Hejazi}},\ }\href {https://doi.org/10.1103/PhysRevFluids.5.084006}
  {\bibfield  {journal} {\bibinfo  {journal} {Phys.~Rev.~Fluids}\ }\textbf
  {\bibinfo {volume} {5}},\ \bibinfo {pages} {084006} (\bibinfo {year}
  {2020})}\BibitemShut {NoStop}%
\bibitem [{\citenamefont {Atasi}\ \emph {et~al.}(2018)\citenamefont {Atasi},
  \citenamefont {Haut}, \citenamefont {Pedrono}, \citenamefont {Scheid},\ and\
  \citenamefont {Legendre}}]{Atasi2018}%
  \BibitemOpen
  \bibfield  {author} {\bibinfo {author} {\bibfnamefont {O.}~\bibnamefont
  {Atasi}}, \bibinfo {author} {\bibfnamefont {B.}~\bibnamefont {Haut}},
  \bibinfo {author} {\bibfnamefont {A.}~\bibnamefont {Pedrono}}, \bibinfo
  {author} {\bibfnamefont {B.}~\bibnamefont {Scheid}},\ and\ \bibinfo {author}
  {\bibfnamefont {D.}~\bibnamefont {Legendre}},\ }\href
  {https://doi.org/10.1021/acs.langmuir.8b01805} {\bibfield  {journal}
  {\bibinfo  {journal} {Langmuir}\ }\textbf {\bibinfo {volume} {34}},\ \bibinfo
  {pages} {10048} (\bibinfo {year} {2018})}\BibitemShut {NoStop}%
\bibitem [{\citenamefont {Walstra}(1993)}]{Walstra1993}%
  \BibitemOpen
  \bibfield  {author} {\bibinfo {author} {\bibfnamefont {P.}~\bibnamefont
  {Walstra}},\ }\href {https://doi.org/10.1016/0009-2509(93)80021-H} {\bibfield
   {journal} {\bibinfo  {journal} {Chem.~Eng.~Sci.}\ }\textbf {\bibinfo
  {volume} {48}},\ \bibinfo {pages} {333} (\bibinfo {year} {1993})}\BibitemShut
  {NoStop}%
\bibitem [{\citenamefont {Firouzi}\ and\ \citenamefont
  {Nguyen}(2017)}]{Firouzi2017}%
  \BibitemOpen
  \bibfield  {author} {\bibinfo {author} {\bibfnamefont {M.}~\bibnamefont
  {Firouzi}}\ and\ \bibinfo {author} {\bibfnamefont {A.~V.}\ \bibnamefont
  {Nguyen}},\ }\href {https://doi.org/10.1016/j.colsurfa.2016.12.004}
  {\bibfield  {journal} {\bibinfo  {journal} {Colloids and Surfaces A}\
  }\textbf {\bibinfo {volume} {515}},\ \bibinfo {pages} {62} (\bibinfo {year}
  {2017})}\BibitemShut {NoStop}%
\bibitem [{\citenamefont {Espitia}\ \emph {et~al.}(2019)\citenamefont
  {Espitia}, \citenamefont {Fuenmayor},\ and\ \citenamefont
  {Otoni}}]{Espitia2019}%
  \BibitemOpen
  \bibfield  {author} {\bibinfo {author} {\bibfnamefont {P.~J.~P.}\
  \bibnamefont {Espitia}}, \bibinfo {author} {\bibfnamefont {C.~A.}\
  \bibnamefont {Fuenmayor}},\ and\ \bibinfo {author} {\bibfnamefont {C.~G.}\
  \bibnamefont {Otoni}},\ }\href {https://doi.org/10.1111/1541-4337.12405}
  {\bibfield  {journal} {\bibinfo  {journal} {Comprehensive Reviews in Food
  Science and Food Safety}\ }\textbf {\bibinfo {volume} {18}},\ \bibinfo
  {pages} {264} (\bibinfo {year} {2019})}\BibitemShut {NoStop}%
\bibitem [{\citenamefont {Hack}\ \emph {et~al.}(2021)\citenamefont {Hack},
  \citenamefont {Kwieci{\'{n}}ski}, \citenamefont {Ram{\'{i}}rez-Soto},
  \citenamefont {Segers}, \citenamefont {Karpitschka}, \citenamefont {Kooij},\
  and\ \citenamefont {Snoeijer}}]{Hack2021}%
  \BibitemOpen
  \bibfield  {author} {\bibinfo {author} {\bibfnamefont {M.~A.}\ \bibnamefont
  {Hack}}, \bibinfo {author} {\bibfnamefont {W.}~\bibnamefont
  {Kwieci{\'{n}}ski}}, \bibinfo {author} {\bibfnamefont {O.}~\bibnamefont
  {Ram{\'{i}}rez-Soto}}, \bibinfo {author} {\bibfnamefont {T.}~\bibnamefont
  {Segers}}, \bibinfo {author} {\bibfnamefont {S.}~\bibnamefont {Karpitschka}},
  \bibinfo {author} {\bibfnamefont {E.~S.}\ \bibnamefont {Kooij}},\ and\
  \bibinfo {author} {\bibfnamefont {J.~H.}\ \bibnamefont {Snoeijer}},\ }\href
  {https://doi.org/10.1021/acs.langmuir.0c03571} {\bibfield  {journal}
  {\bibinfo  {journal} {Langmuir}\ }\textbf {\bibinfo {volume} {37}},\ \bibinfo
  {pages} {3605} (\bibinfo {year} {2021})}\BibitemShut {NoStop}%
\bibitem [{\citenamefont {Eijkel}\ and\ \citenamefont {van~den
  Berg}(2005)}]{Eijkel2005}%
  \BibitemOpen
  \bibfield  {author} {\bibinfo {author} {\bibfnamefont {J.~C.}\ \bibnamefont
  {Eijkel}}\ and\ \bibinfo {author} {\bibfnamefont {A.}~\bibnamefont {van~den
  Berg}},\ }\href {https://doi.org/10.1007/s10404-004-0012-9} {\bibfield
  {journal} {\bibinfo  {journal} {Microfluidics and Nanofluidics}\ }\textbf
  {\bibinfo {volume} {1}},\ \bibinfo {pages} {249} (\bibinfo {year}
  {2005})}\BibitemShut {NoStop}%
\bibitem [{\citenamefont {Bocquet}\ and\ \citenamefont
  {Charlaix}(2010)}]{Bocquet2010}%
  \BibitemOpen
  \bibfield  {author} {\bibinfo {author} {\bibfnamefont {L.}~\bibnamefont
  {Bocquet}}\ and\ \bibinfo {author} {\bibfnamefont {E.}~\bibnamefont
  {Charlaix}},\ }\href {https://doi.org/10.1039/B909366B} {\bibfield  {journal}
  {\bibinfo  {journal} {Chem. Soc. Rev.}\ }\textbf {\bibinfo {volume} {39}},\
  \bibinfo {pages} {1073} (\bibinfo {year} {2010})}\BibitemShut {NoStop}%
\bibitem [{\citenamefont {Imai}\ \emph {et~al.}(2017)\citenamefont {Imai},
  \citenamefont {Yamamoto}, \citenamefont {Okano}, \citenamefont {Sato},\ and\
  \citenamefont {Shigeta}}]{Imai2017}%
  \BibitemOpen
  \bibfield  {author} {\bibinfo {author} {\bibfnamefont {Y.}~\bibnamefont
  {Imai}}, \bibinfo {author} {\bibfnamefont {T.}~\bibnamefont {Yamamoto}},
  \bibinfo {author} {\bibfnamefont {Y.}~\bibnamefont {Okano}}, \bibinfo
  {author} {\bibfnamefont {R.}~\bibnamefont {Sato}},\ and\ \bibinfo {author}
  {\bibfnamefont {Y.}~\bibnamefont {Shigeta}},\ }\href@noop {} {\bibfield
  {journal} {\bibinfo  {journal} {ASEAN J.~Chem.~Eng.}\ }\textbf {\bibinfo
  {volume} {17}},\ \bibinfo {pages} {29} (\bibinfo {year} {2017})}\BibitemShut
  {NoStop}%
\bibitem [{\citenamefont {Imai}\ \emph {et~al.}(2019)\citenamefont {Imai},
  \citenamefont {Yamamoto}, \citenamefont {Sekimoto}, \citenamefont {Okano},
  \citenamefont {Sato},\ and\ \citenamefont {Shigeta}}]{Imai2019}%
  \BibitemOpen
  \bibfield  {author} {\bibinfo {author} {\bibfnamefont {Y.}~\bibnamefont
  {Imai}}, \bibinfo {author} {\bibfnamefont {T.}~\bibnamefont {Yamamoto}},
  \bibinfo {author} {\bibfnamefont {A.}~\bibnamefont {Sekimoto}}, \bibinfo
  {author} {\bibfnamefont {Y.}~\bibnamefont {Okano}}, \bibinfo {author}
  {\bibfnamefont {R.}~\bibnamefont {Sato}},\ and\ \bibinfo {author}
  {\bibfnamefont {Y.}~\bibnamefont {Shigeta}},\ }\href
  {https://doi.org/10.1016/j.jtice.2018.05.015} {\bibfield  {journal} {\bibinfo
   {journal} {J.~Taiwan Inst.~Chem.~Eng.}\ }\textbf {\bibinfo {volume} {98}},\
  \bibinfo {pages} {20} (\bibinfo {year} {2019})}\BibitemShut {NoStop}%
\bibitem [{\citenamefont {Liu}\ \emph {et~al.}(2017)\citenamefont {Liu},
  \citenamefont {Ganti}, \citenamefont {Burton}, \citenamefont {Zhang},
  \citenamefont {Wang},\ and\ \citenamefont {Frenkel}}]{Liu2017}%
  \BibitemOpen
  \bibfield  {author} {\bibinfo {author} {\bibfnamefont {Y.}~\bibnamefont
  {Liu}}, \bibinfo {author} {\bibfnamefont {R.}~\bibnamefont {Ganti}}, \bibinfo
  {author} {\bibfnamefont {H.~G.}\ \bibnamefont {Burton}}, \bibinfo {author}
  {\bibfnamefont {X.}~\bibnamefont {Zhang}}, \bibinfo {author} {\bibfnamefont
  {W.}~\bibnamefont {Wang}},\ and\ \bibinfo {author} {\bibfnamefont
  {D.}~\bibnamefont {Frenkel}},\ }\href
  {https://doi.org/10.1103/PhysRevLett.119.224502} {\bibfield  {journal}
  {\bibinfo  {journal} {Phys.~Rev.~Lett.}\ }\textbf {\bibinfo {volume} {119}},\
  \bibinfo {pages} {224502} (\bibinfo {year} {2017})}\BibitemShut {NoStop}%
\bibitem [{\citenamefont {Galliéro}(2010)}]{Galliero2010}%
  \BibitemOpen
  \bibfield  {author} {\bibinfo {author} {\bibfnamefont {G.}~\bibnamefont
  {Galliéro}},\ }\href {https://doi.org/10.1103/PhysRevE.81.056306} {\bibfield
   {journal} {\bibinfo  {journal} {Phys.~Rev.~E}\ }\textbf {\bibinfo {volume}
  {81}},\ \bibinfo {pages} {056306} (\bibinfo {year} {2010})}\BibitemShut
  {NoStop}%
\bibitem [{\citenamefont {Poesio}\ \emph {et~al.}(2017)\citenamefont {Poesio},
  \citenamefont {Damone},\ and\ \citenamefont {Matar}}]{Poesio2017}%
  \BibitemOpen
  \bibfield  {author} {\bibinfo {author} {\bibfnamefont {P.}~\bibnamefont
  {Poesio}}, \bibinfo {author} {\bibfnamefont {A.}~\bibnamefont {Damone}},\
  and\ \bibinfo {author} {\bibfnamefont {O.~K.}\ \bibnamefont {Matar}},\ }\href
  {https://doi.org/10.1103/PhysRevFluids.2.044004} {\bibfield  {journal}
  {\bibinfo  {journal} {Phys.~Rev.~Fluids}\ }\textbf {\bibinfo {volume} {2}},\
  \bibinfo {pages} {044004} (\bibinfo {year} {2017})}\BibitemShut {NoStop}%
\bibitem [{\citenamefont {Zhan}\ \emph {et~al.}(2020)\citenamefont {Zhan},
  \citenamefont {Su}, \citenamefont {Jin}, \citenamefont {Zhang}, \citenamefont
  {Wang}, \citenamefont {Hao},\ and\ \citenamefont {Li}}]{Zhan2020}%
  \BibitemOpen
  \bibfield  {author} {\bibinfo {author} {\bibfnamefont {S.}~\bibnamefont
  {Zhan}}, \bibinfo {author} {\bibfnamefont {Y.}~\bibnamefont {Su}}, \bibinfo
  {author} {\bibfnamefont {Z.}~\bibnamefont {Jin}}, \bibinfo {author}
  {\bibfnamefont {M.}~\bibnamefont {Zhang}}, \bibinfo {author} {\bibfnamefont
  {W.}~\bibnamefont {Wang}}, \bibinfo {author} {\bibfnamefont {Y.}~\bibnamefont
  {Hao}},\ and\ \bibinfo {author} {\bibfnamefont {L.}~\bibnamefont {Li}},\
  }\href {https://doi.org/10.1016/j.cej.2020.125053} {\bibfield  {journal}
  {\bibinfo  {journal} {Chem.~Eng.~J.}\ }\textbf {\bibinfo {volume} {395}},\
  \bibinfo {pages} {125053} (\bibinfo {year} {2020})}\BibitemShut {NoStop}%
\bibitem [{\citenamefont {Humphrey}\ \emph {et~al.}(1996)\citenamefont
  {Humphrey}, \citenamefont {Dalke},\ and\ \citenamefont
  {Schulten}}]{Humphrey1996}%
  \BibitemOpen
  \bibfield  {author} {\bibinfo {author} {\bibfnamefont {W.}~\bibnamefont
  {Humphrey}}, \bibinfo {author} {\bibfnamefont {A.}~\bibnamefont {Dalke}},\
  and\ \bibinfo {author} {\bibfnamefont {K.}~\bibnamefont {Schulten}},\ }\href
  {https://doi.org/10.1016/0263-7855(96)00018-5} {\bibfield  {journal}
  {\bibinfo  {journal} {J.~Mol.~Graphics}\ }\textbf {\bibinfo {volume} {14}},\
  \bibinfo {pages} {33} (\bibinfo {year} {1996})}\BibitemShut {NoStop}%
\bibitem [{\citenamefont {Hunter}(2007)}]{Hunter2007}%
  \BibitemOpen
  \bibfield  {author} {\bibinfo {author} {\bibfnamefont {J.~D.}\ \bibnamefont
  {Hunter}},\ }\href {https://doi.org/10.1109/MCSE.2007.55} {\bibfield
  {journal} {\bibinfo  {journal} {Computing in Sci. \& Eng.}\ }\textbf
  {\bibinfo {volume} {9}},\ \bibinfo {pages} {90} (\bibinfo {year}
  {2007})}\BibitemShut {NoStop}%
\bibitem [{\citenamefont {Feuersänger}(2021)}]{Feuersanger2021}%
  \BibitemOpen
  \bibfield  {author} {\bibinfo {author} {\bibfnamefont {C.}~\bibnamefont
  {Feuersänger}},\ }\href@noop {} {\bibinfo {title} {{PGFPlots -- A LaTeX
  Package to create normal/logarithmic plots in two and three dimensions}}},\
  \bibinfo {howpublished} {available at \url{http://pgfplots.sourceforge.net/}}
  (\bibinfo {year} {2021})\BibitemShut {NoStop}%
\bibitem [{\citenamefont {Allen}\ and\ \citenamefont
  {Tildesley}(2017)}]{AllenTildesley2017}%
  \BibitemOpen
  \bibfield  {author} {\bibinfo {author} {\bibfnamefont {M.}~\bibnamefont
  {Allen}}\ and\ \bibinfo {author} {\bibfnamefont {D.}~\bibnamefont
  {Tildesley}},\ }\href@noop {} {\emph {\bibinfo {title} {Computer Simulation
  of Liquids}}}\ (\bibinfo  {publisher} {OUP Oxford},\ \bibinfo {year}
  {2017})\BibitemShut {NoStop}%
\bibitem [{\citenamefont {Bordat}\ and\ \citenamefont
  {M{\"{u}}ller-Plathe}(2002)}]{Bordat2002}%
  \BibitemOpen
  \bibfield  {author} {\bibinfo {author} {\bibfnamefont {P.}~\bibnamefont
  {Bordat}}\ and\ \bibinfo {author} {\bibfnamefont {F.}~\bibnamefont
  {M{\"{u}}ller-Plathe}},\ }\href {https://doi.org/10.1063/1.1436124}
  {\bibfield  {journal} {\bibinfo  {journal} {J.~Chem.~Phys.}\ }\textbf
  {\bibinfo {volume} {116}},\ \bibinfo {pages} {3362} (\bibinfo {year}
  {2002})}\BibitemShut {NoStop}%
\bibitem [{\citenamefont {Galliéro}\ \emph {et~al.}(2005)\citenamefont
  {Galliéro}, \citenamefont {Boned},\ and\ \citenamefont
  {Baylaucq}}]{Galliero2005}%
  \BibitemOpen
  \bibfield  {author} {\bibinfo {author} {\bibfnamefont {G.}~\bibnamefont
  {Galliéro}}, \bibinfo {author} {\bibfnamefont {C.}~\bibnamefont {Boned}},\
  and\ \bibinfo {author} {\bibfnamefont {A.}~\bibnamefont {Baylaucq}},\ }\href
  {https://doi.org/10.1021/ie050154t} {\bibfield  {journal} {\bibinfo
  {journal} {Ind.~Eng.~Chem.~Res.}\ }\textbf {\bibinfo {volume} {44}},\
  \bibinfo {pages} {6963} (\bibinfo {year} {2005})}\BibitemShut {NoStop}%
\bibitem [{\citenamefont {Galliéro}\ and\ \citenamefont
  {Boned}(2009)}]{Galliero2009}%
  \BibitemOpen
  \bibfield  {author} {\bibinfo {author} {\bibfnamefont {G.}~\bibnamefont
  {Galliéro}}\ and\ \bibinfo {author} {\bibfnamefont {C.}~\bibnamefont
  {Boned}},\ }\href {https://doi.org/10.1103/PhysRevE.79.021201} {\bibfield
  {journal} {\bibinfo  {journal} {Phys.~Rev.~E}\ }\textbf {\bibinfo {volume}
  {79}},\ \bibinfo {pages} {021201} (\bibinfo {year} {2009})}\BibitemShut
  {NoStop}%
\bibitem [{\citenamefont {Abraham}\ \emph {et~al.}(2015)\citenamefont
  {Abraham}, \citenamefont {Murtola}, \citenamefont {Schulz}, \citenamefont
  {P{\'{a}}ll}, \citenamefont {Smith}, \citenamefont {Hess},\ and\
  \citenamefont {Lindahl}}]{Abraham2015}%
  \BibitemOpen
  \bibfield  {author} {\bibinfo {author} {\bibfnamefont {M.~J.}\ \bibnamefont
  {Abraham}}, \bibinfo {author} {\bibfnamefont {T.}~\bibnamefont {Murtola}},
  \bibinfo {author} {\bibfnamefont {R.}~\bibnamefont {Schulz}}, \bibinfo
  {author} {\bibfnamefont {S.}~\bibnamefont {P{\'{a}}ll}}, \bibinfo {author}
  {\bibfnamefont {J.~C.}\ \bibnamefont {Smith}}, \bibinfo {author}
  {\bibfnamefont {B.}~\bibnamefont {Hess}},\ and\ \bibinfo {author}
  {\bibfnamefont {E.}~\bibnamefont {Lindahl}},\ }\href
  {https://doi.org/10.1016/j.softx.2015.06.001} {\bibfield  {journal} {\bibinfo
   {journal} {SoftwareX}\ }\textbf {\bibinfo {volume} {1-2}},\ \bibinfo {pages}
  {19} (\bibinfo {year} {2015})}\BibitemShut {NoStop}%
\bibitem [{\citenamefont {P{\'{a}}ll}\ \emph {et~al.}(2020)\citenamefont
  {P{\'{a}}ll}, \citenamefont {Zhmurov}, \citenamefont {Bauer}, \citenamefont
  {Abraham}, \citenamefont {Lundborg}, \citenamefont {Gray}, \citenamefont
  {Hess},\ and\ \citenamefont {Lindahl}}]{Pall2020}%
  \BibitemOpen
  \bibfield  {author} {\bibinfo {author} {\bibfnamefont {S.}~\bibnamefont
  {P{\'{a}}ll}}, \bibinfo {author} {\bibfnamefont {A.}~\bibnamefont {Zhmurov}},
  \bibinfo {author} {\bibfnamefont {P.}~\bibnamefont {Bauer}}, \bibinfo
  {author} {\bibfnamefont {M.~J.}\ \bibnamefont {Abraham}}, \bibinfo {author}
  {\bibfnamefont {M.}~\bibnamefont {Lundborg}}, \bibinfo {author}
  {\bibfnamefont {A.}~\bibnamefont {Gray}}, \bibinfo {author} {\bibfnamefont
  {B.}~\bibnamefont {Hess}},\ and\ \bibinfo {author} {\bibfnamefont
  {E.}~\bibnamefont {Lindahl}},\ }\href {https://doi.org/10.1063/5.0018516}
  {\bibfield  {journal} {\bibinfo  {journal} {J.~Chem.~Phys.}\ }\textbf
  {\bibinfo {volume} {153}},\ \bibinfo {pages} {134110} (\bibinfo {year}
  {2020})}\BibitemShut {NoStop}%
\bibitem [{\citenamefont {Bussi}\ \emph {et~al.}(2007)\citenamefont {Bussi},
  \citenamefont {Donadio},\ and\ \citenamefont {Parrinello}}]{Bussi2007}%
  \BibitemOpen
  \bibfield  {author} {\bibinfo {author} {\bibfnamefont {G.}~\bibnamefont
  {Bussi}}, \bibinfo {author} {\bibfnamefont {D.}~\bibnamefont {Donadio}},\
  and\ \bibinfo {author} {\bibfnamefont {M.}~\bibnamefont {Parrinello}},\
  }\href {https://doi.org/10.1063/1.2408420} {\bibfield  {journal} {\bibinfo
  {journal} {J.~Chem.~Phys.}\ }\textbf {\bibinfo {volume} {126}},\ \bibinfo
  {pages} {014101} (\bibinfo {year} {2007})}\BibitemShut {NoStop}%
\bibitem [{Sup()}]{SupplementaryMaterial}%
  \BibitemOpen
  \href@noop {} {}\bibinfo {note} {See Supplementary Materials at [URL] for
  additional figures and a derivation of
  \eqref{eq:marangoni-flow-velocity}.}\BibitemShut {Stop}%
\end{thebibliography}
\end{document}


\title{Supplementary Material -- How molecular effects affect solutal Marangoni convection}

\author{Petter Johansson}
\author{Guillaume Galliéro}
\affiliation{Laboratoire des Fluides Complexes et Leurs Réservoirs, Université de Pau et des Pays de l’Adour, E2S UPPA, CNRS, Total, LFCR, Pau, France}

\author{Dominique Legendre}
\affiliation{Institut de Mécanique des Fluides de Toulouse (IMFT), Université de Toulouse, CNRS-INPT-UPS, Toulouse, France}

\date{\today}

\maketitle

\section{Deriving equation 4:\\ Interfacial Marangoni flow}
\noindent

This is a derivation of equation 4 in the main paper: the expected Marangoni flow velocity $u_{x,I}$ at the interfaces for our two-phase systems. 

\subsection{Assumptions}

Consider a two-dimensional two-phase system contained in a box of size $L \times 2H$ along $x$ and $z$, with an interface between the phases at $z = H$.  This mimics a single ``quarter'' zone of our simulated systems. See figure below. In addition, the system is mirrored around $z = 0$.

\begin{center}
    \includegraphics{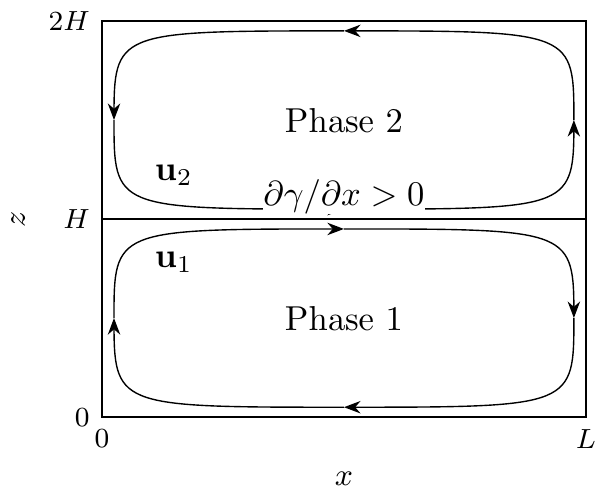}
\end{center}

The two phases are incompressible and immiscible but have a common density $\rho$ and viscosity $\mu$. At the interface we have a gradient of interfacial tension $\di{\gamma}/\dx$. We consider the limit where the interface width is negligible compared to the system size. Finally, we consider the case when a steady state has formed.

We now set up the following boundary conditions:

\begin{enumerate}
    \item Far away from the domain boundaries the flow is parallel to the interface:
        \begin{equation}
            \label{bc:parallel-flow}
            \myvec{u}_{k}(x, z) = u_{x,k}(x, z) \unitvec{x}
        \end{equation}
        where $k = 1,2$ is the phase index. Since the flow is incompressible we have in this region:
        \begin{align}
            \nabla \cdot \myvec{u}_{k} = {}& 0 \nonumber \\
            \label{bc:flow-constant-along-x}
            \Leftrightarrow \quad {}& \ppx{u_{x,k}} = 0 \\
            \Rightarrow \quad {}& u_{x,k} = u_{x,k}(z)
        \end{align}

    \item At $z = 0$ the flow gradient is smooth and continuous (the mirror symmetry introduces the same phase on the other side):
        \begin{equation}
            \label{bc:flow-smooth-at-zero}
            \left . \ppz{u_{x,1}} \right \rvert_{z = 0} = 0
        \end{equation}

    \item At $z = H$ the flow of the two phases is continuous and mirrored around the interface:
        \begin{align}
            \label{bc:flow-continuous-at-interface}
            u_{x,1}(z = H^-) = {}& u_{x,2}(z = H^+) \equiv u_{x,I} \\
            \label{bc:flow-mirrored-at-interface}
            \left . \ppz{u_{x,1}} \right \rvert_{z = H^-} 
                = {}& \left . -\ppz{u_{x,2}} \right \rvert_{z = H^+}
        \end{align}
    
    \item Since mass is conserved there is no net flow along $x$ through a phase:
        \begin{equation}
            \label{bc:zero-net-flow}
            \int_0^H \dz \, u_{x,k}(z) = 0 \, .
        \end{equation}
\end{enumerate}

\subsection{Momentum conservation at interface}
For a planar interface, i.e. without curvature, momentum conservation at the interface gives 
\begin{equation}
    \label{eq:momentum-conservation-setup}
    -\nabla \gamma = \sum_{k=1,2} \parens{ 
        p_k \mytensor{I} - \mytensor{\tau}_k
    } \cdot \unitvec{n}_k
\end{equation}
where $\mytensor{\tau}_k$ is the stress tensor, $p_k$ the pressure and $\unitvec{n}_k$ the interface normal for phase $k$. $\mytensor{I}$ is the identity matrix. The normals are $\unitvec{n}_1 = \unitvec{z}$ and $\unitvec{n}_2 = -\unitvec{z}$.

With \eqref{bc:parallel-flow} and \eqref{bc:flow-constant-along-x} the stress tensor resolves into 
\begin{equation}
    \begin{split}
        \mytensor{\tau}_k
            = {}& \mu \begin{bmatrix}
                2 \ppx{u_{x,k}} & \ppx{u_{z,k}} + \ppz{u_{x,k}} \\
                \ppx{u_{z,k}} + \ppz{u_{x,k}} & 2 \ppz{u_{z,k}}
            \end{bmatrix} \\
            = {}& \mu \begin{bmatrix}
                0 & \ppz{u_{x,k}} \\
                \ppz{u_{x,k}} & 0
            \end{bmatrix} \, .
    \end{split} 
\end{equation}
Together with \eqref{bc:flow-mirrored-at-interface} we can now solve \eqref{eq:momentum-conservation-setup}:
\begin{equation}
    \begin{split}
        \begin{bmatrix}
            -\ddx{\gamma} \\ 
            0
        \end{bmatrix} 
    = {}& \begin{bmatrix}
            \mu \parens{\ppz{u_{x,2}} - \ppz{u_{x,1}}} \\
            p_1 - p_2
        \end{bmatrix} \\
    = {}& \begin{bmatrix}
            -2\mu \ppz{u_{x,1}} \\
            p_1 - p_2
        \end{bmatrix} 
    \end{split}
\end{equation}
and extract 
\begin{equation}
    \label{eq:momentum-conservation}
    \left. \ppz{u_{x,1}} \right \rvert_{z = H} = \frac{1}{2\mu} \ddx{\gamma} \, .
\end{equation}

\subsection{Calculation}

We set up the incompressible Navier--Stokes equation inside the liquid:
\begin{equation}
    \label{eq:navier-stokes}
    \rho \parens{
        \deriv{\myvec{u}}{t}
        + \parens{\myvec{u} \cdot \nabla} \myvec{u}
        } 
        = -\nabla p + \mu \nabla^2 \myvec{u}
\end{equation}

Since the two phases are mirrored we now consider the flow in phase 1. Since we are in a steady state we have
\begin{equation}
    \ddt{\myvec{u}_1} = 0 \, . 
\end{equation}
With \eqref{bc:parallel-flow} and \eqref{bc:flow-constant-along-x}:
\begin{equation}
    \begin{split}
    \parens{\myvec{u}_1 \cdot \nabla} \myvec{u}_1 = {}& u_{x,1} \ppx{u_{x,1}} \\
        = {}& 0 \, .
    \end{split}
\end{equation}
This allows us to rewrite \eqref{eq:navier-stokes} as:

\begin{align}
    \rho \deriv{\myvec{u}_1}{t}
        + \rho \parens{\myvec{u}_1 \cdot \nabla} \myvec{u}_1
    = {}& -\nabla p_1 + \mu \nabla^2 \myvec{u}_1 \\
    \label{eq:navier-stokes-laplacian}
    \Rightarrow \quad \mu \pppz{u_{x,1}} = {}& \ddx{p_1}
\end{align}

We now integrate this term along $z$:
\begin{align}
    \label{eq:second-order-derivative}
    \pppz{u_{x,1}} = {}& \frac{1}{\mu} \ddx{p_1} \equiv A \\
    \label{eq:first-order-derivative}
    \ppz{u_{x,1}} = {}& Az + B \\
    \label{eq:flow-unsolved}
    u_{x,1}(z) = {}& \frac{Az^2}{2} + Bz + C
\end{align}
We evaluate \eqref{eq:first-order-derivative} at $z = 0$ and use boundary condition \eqref{bc:flow-smooth-at-zero} to calculate $B$:
\begin{equation}
    \left. \ppz{u_{x,1}} \right \rvert_{z=0} = B = 0 \, ,
\end{equation}
then at $z = H$ with \eqref{eq:momentum-conservation} to solve for $A$:
\begin{align}
    \begin{split}
        \left. \ppz{u_{x,1}} \right \rvert_{z = H}
            = {}& AH \\
            = {}& \frac{1}{2\mu} \ddx{\gamma}
    \end{split} \\
    \Leftrightarrow \quad A = {}& \frac{1}{2H\mu} \ddx{\gamma}
\end{align}
We finally integrate \eqref{eq:flow-unsolved} and use \eqref{bc:zero-net-flow} to extract $C$:
\begin{align}
    \begin{split}
        \int_0^H \dz \, u_{x,1}(z) = {}& 0 \\
            = {}& \brackets{\frac{Az^3}{6} + \frac{Bz^2}{2} + Cz}_{z=0}^H \\
            = {}& \frac{AH^3}{6} + CH
    \end{split} \\
    \begin{split}
        \Leftrightarrow \; C = {}& -\frac{A H^2}{6} \\
        = {}& -\frac{H}{12\mu} \ddx{\gamma}
    \end{split}
\end{align}

Thus we have 
\begin{equation}
    \begin{split}
        A = {}& \frac{1}{2H\mu} \ddx{\gamma} \\
        B = {}& 0 \\
        C = {}& -\frac{H}{12\mu} \ddx{\gamma}
    \end{split}
\end{equation}
and
\begin{equation}
    \label{eq:velocity-profile}
    u_{x,1}(z) 
        = \frac{H}{4\mu} \ddx{\gamma} \parens{\frac{z^2}{H^2} - \frac{1}{3}} 
\end{equation}
which we can evaluate at $z = H$ to get the interface velocity:
\begin{equation}
    \label{eq:interface-velocity}
    \boxed{
        u_{x,I} \equiv u_{x,1}(H) = \frac{H}{6\mu} \ddx{\gamma}
    }
\end{equation}

\subsection{Discussion}
The above derivation is made for a system which is large enough that the interface width can be taken as 0. This gives us boundary condition \eqref{bc:parallel-flow} at some point sufficiently far away from the system edges (where the flow shifts from tangential to the interface, to normal to it).

\begin{figure}[h!]
    \centering
    \begin{subfigure}{\columnwidth}
        \includegraphics{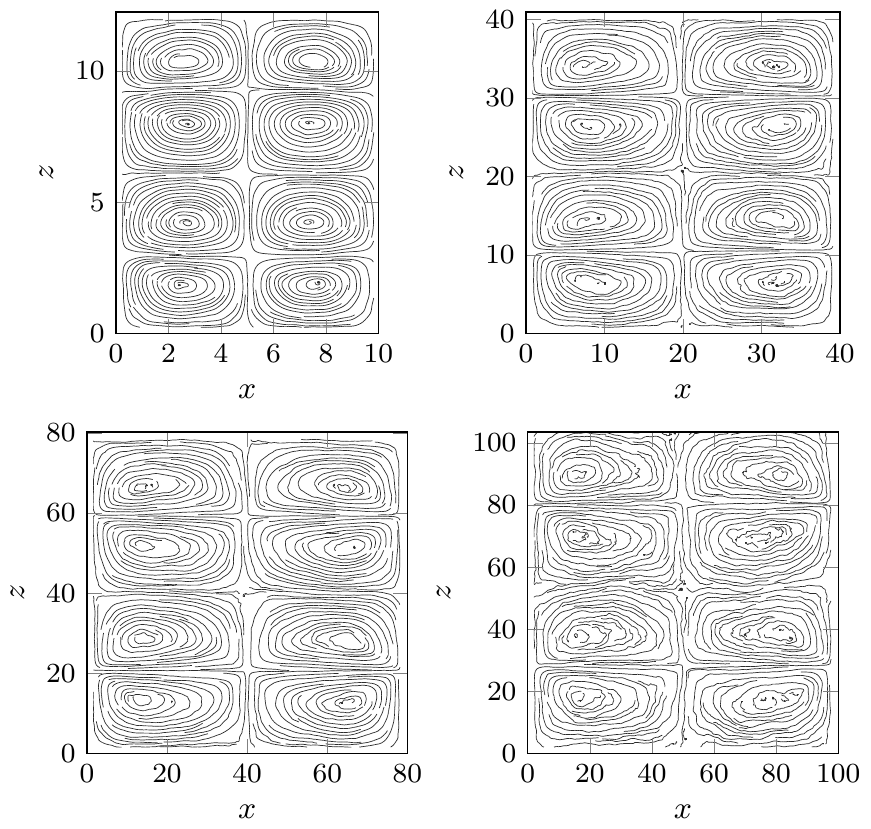}
        \subcaption{1:1 ratio}
    \end{subfigure}
    \newline
    \begin{subfigure}{\columnwidth}
        \includegraphics{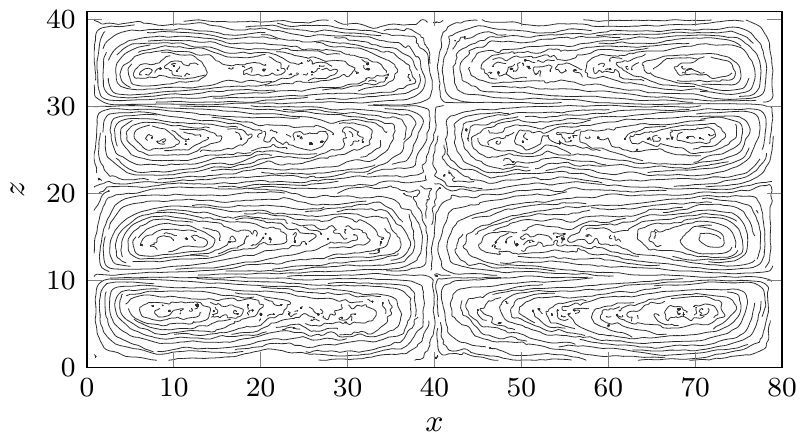}
        \subcaption{2:1 ratio}
    \end{subfigure}
    \caption{\label{fig:flow-field-streamlines} Streamline plots of flow in systems of various sizes.}
\end{figure}

Our systems clearly show finite size effects. Notably, the interface changes width along $x$ since we have a surfactant density gradient. As seen from the streamlines in figure \ref{fig:flow-field-streamlines} the symmetry is disturbed. This affects boundary conditions \eqref{bc:parallel-flow} and \eqref{bc:flow-constant-along-x}. The effect is minimized at points $P$, along the axis of vortices. Accounting for the variable width at these points would be a second order correction.

\clearpage


\begin{figure}[h!]
    \begin{subfigure}{\columnwidth}
        \centering
        \includegraphics{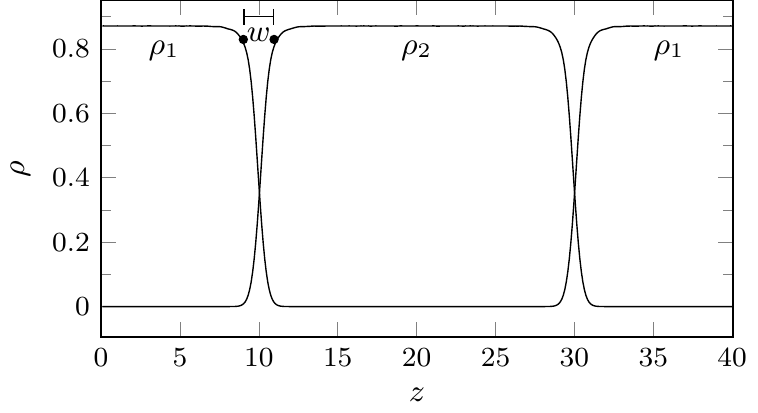}
        \subcaption{$\rho_I = 0$}
    \end{subfigure}
    \newline
    \begin{subfigure}{\columnwidth}
        \centering
        \includegraphics{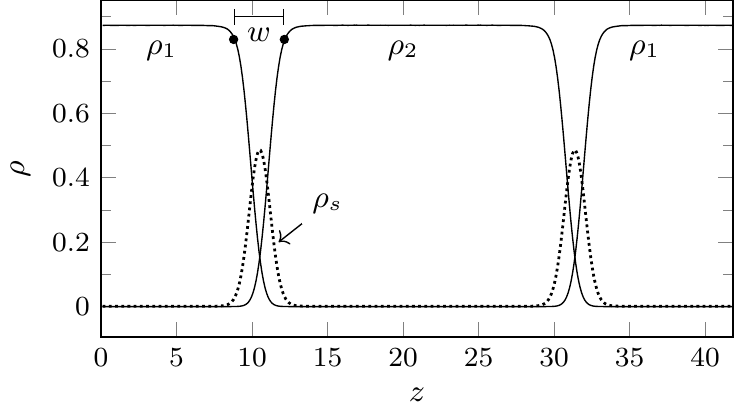}
        \subcaption{$\rho_I = 0.48$}
    \end{subfigure}
    \caption{\label{fig:interface-width} Phase densities $\rho_i$ of the system used to measure interface width $w(\rho_I)$ in, for two different inserted surfactant concentrations $\rho_I$. $\rho_s$ is the surfactant density profile. Bulk phases are defined as $\rho_i \geq 0.95 \rho_b$ where $\rho_b$ is the bulk density.}
\end{figure}


\begin{figure}[h!]
    \centering
    \begin{subfigure}{\columnwidth}
        \centering
        \includegraphics{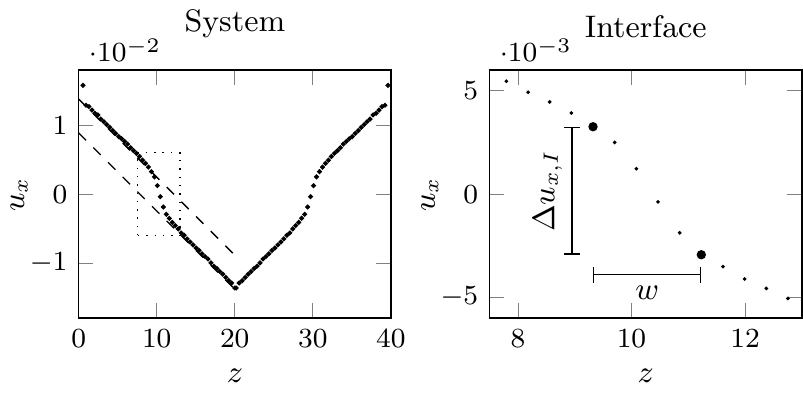}
        \subcaption{$\rho_I = 0$}
    \end{subfigure}
    \newline
    \begin{subfigure}{\columnwidth}
        \centering
        \includegraphics{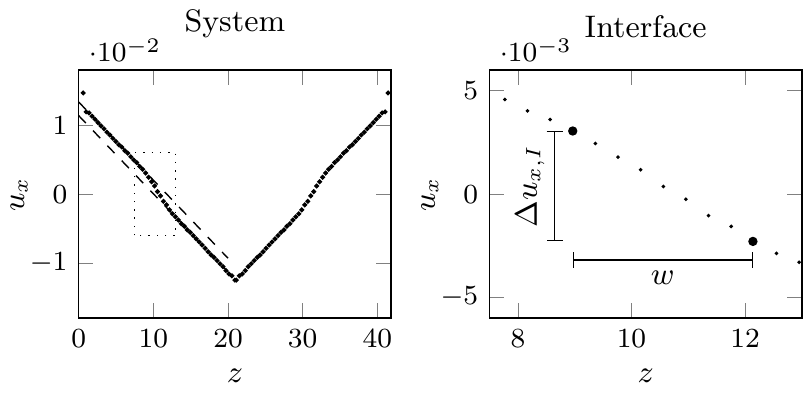}
        \subcaption{$\rho_I = 0.48$}
    \end{subfigure}
    \caption{\label{fig:slip-velocity-profile} Velocity $u_x$ throughout two-phase systems with applied shear $\tau_{xz}$ between the edges and center, for two different surfactant concentrations $\rho_I$. The inset shows the velocity jump $\Delta u_{x,I}(\rho_I)$ across the interface of width $w(\rho_I)$ (as defined in figure \ref{fig:interface-width}).}
\end{figure}


\begin{figure}[h!]
    %
    %
    %
    \centering
    \begin{subfigure}{\columnwidth}
        \centering
        \includegraphics{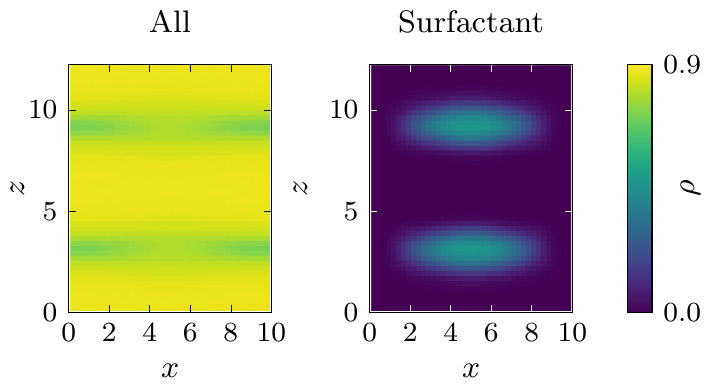}
        \subcaption{$10 \times 10$}
    \end{subfigure}
    \newline
    \begin{subfigure}{\columnwidth}
        \includegraphics{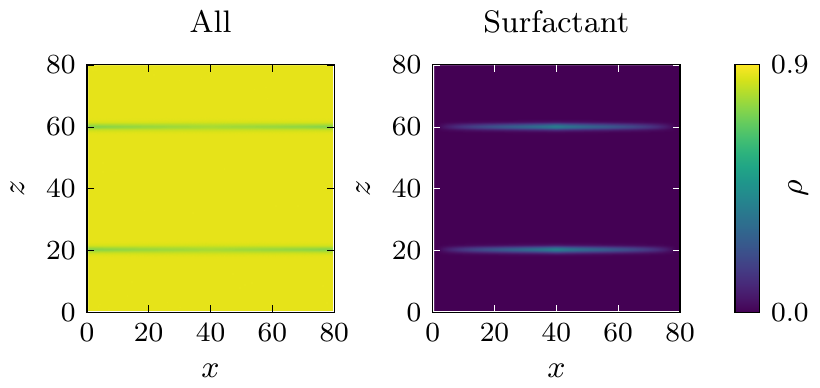}
        \subcaption{$80 \times 80$}
    \end{subfigure}
    \caption{\label{fig:density-plots} Mass density $\rho$ plots of simulated systems in their steady state. On the left is the entire system density, on the right only the surfactant density.}
\end{figure}


\begin{figure}[h!]
    \centering
    \begin{subfigure}{\columnwidth}
        \centering
        \includegraphics{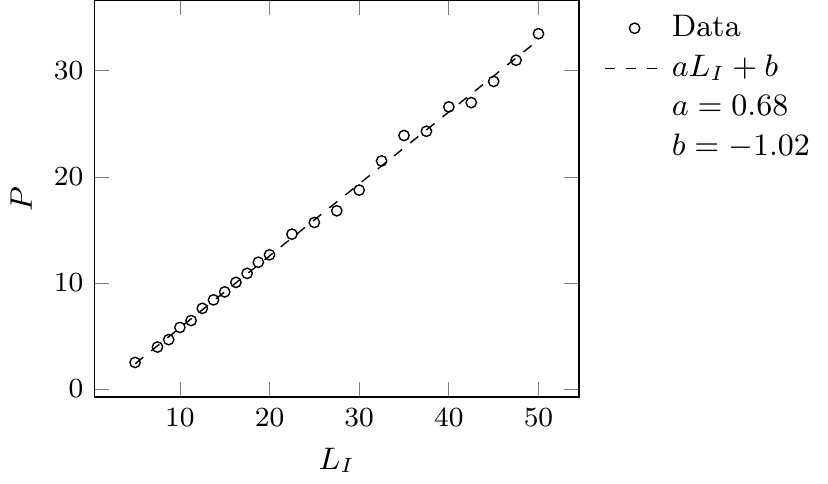}
        \subcaption{}
    \end{subfigure}
    \newline
    \begin{subfigure}{\columnwidth}
        \centering
        \includegraphics{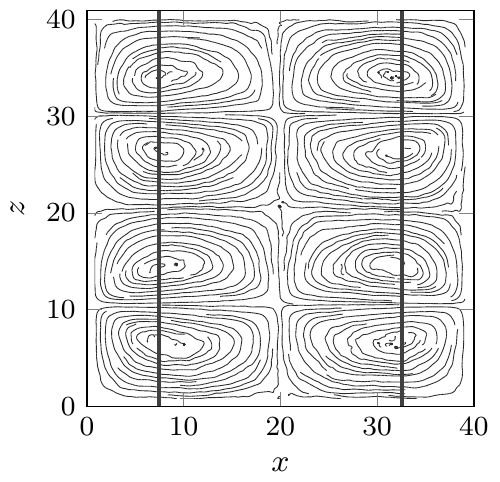}
        \subcaption{}
    \end{subfigure}
    \caption{\label{fig:flow-field-centers} (a) Flow field vortex center positions $P$ along each interface of length $L_I = L_x / 2$. Defines the reference axis through vortex centers for each system size. (b) Corresponding axis along $z$ marked in a $40 \times 40$ system.}
\end{figure}



\begin{figure*}
    \centering
    \includegraphics{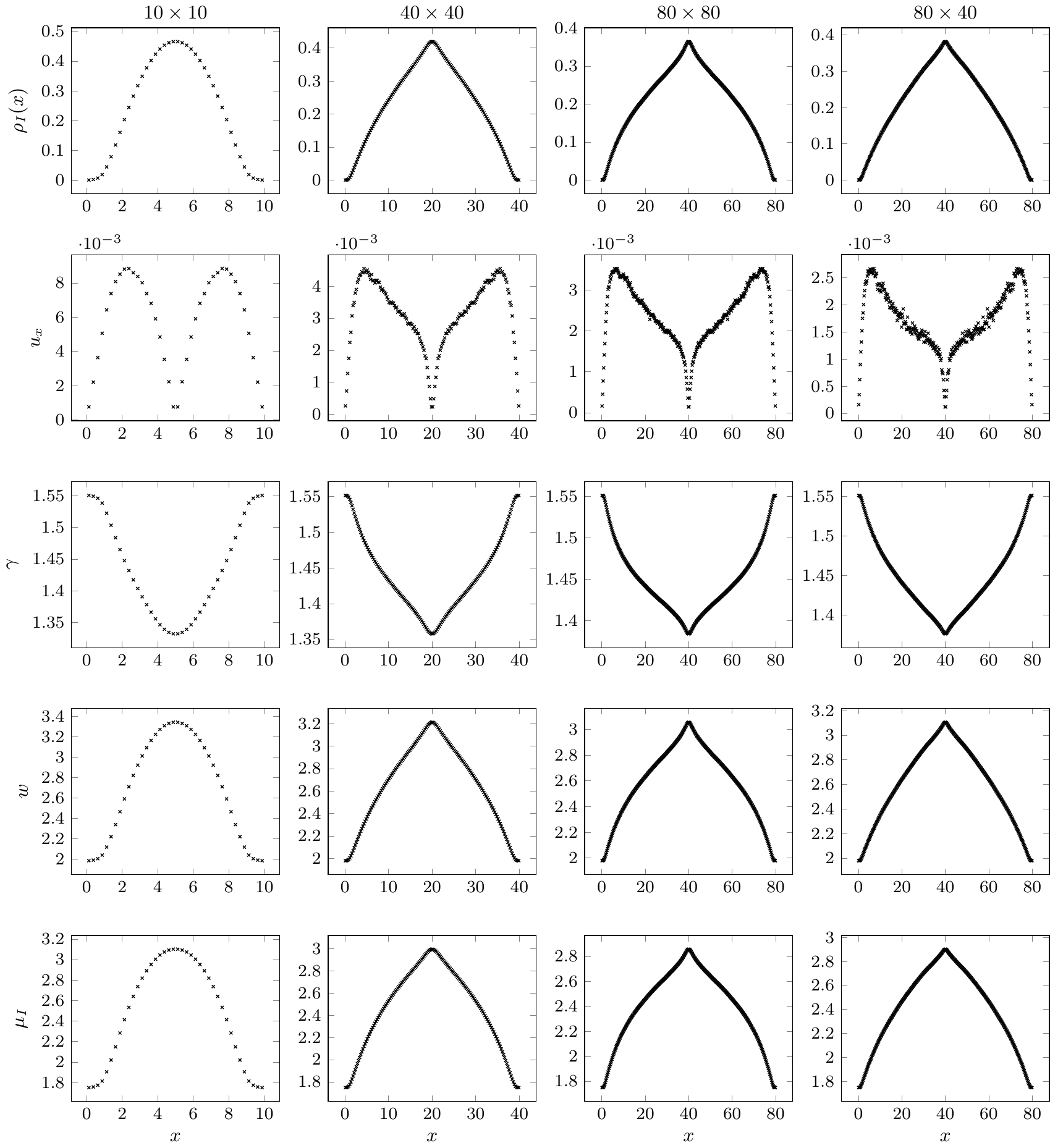}
    \caption{\label{fig:all-parameters-along-interface} Surfactant concentration $\rho_I$ and flow velocity $u_x$ along the interface for a set of system sizes, along with all properties (interfacial tension $\gamma$, interface width $w$, interfacial viscosity $\mu_I$) computed from $\rho_I$ at the same points.}
\end{figure*}


\begin{figure}[t]
    \centering
    \begin{subfigure}{\columnwidth}
        \includegraphics{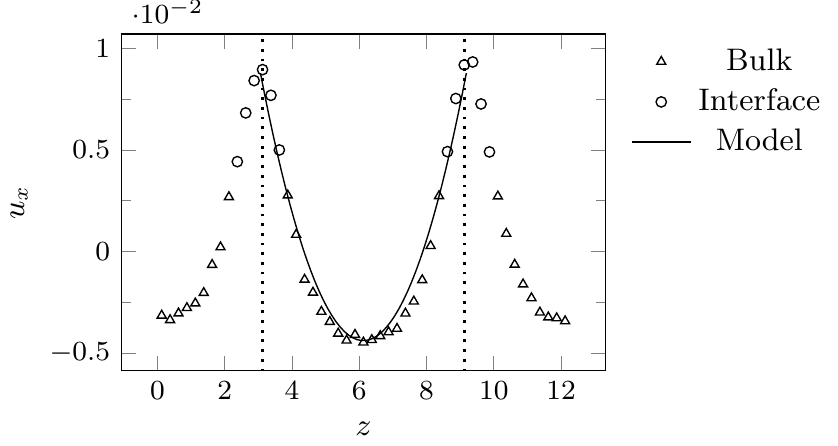}
        \subcaption{$10 \times 10$}
    \end{subfigure}
    \newline
    \begin{subfigure}{\columnwidth}
        \includegraphics{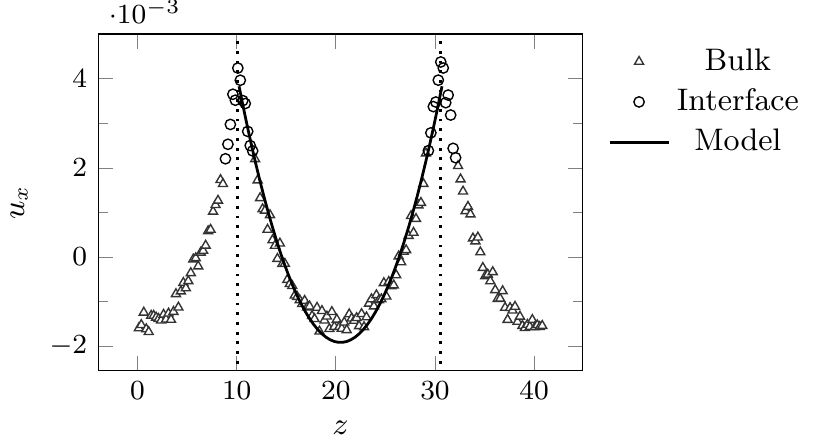}
        \subcaption{$40 \times 40$}
    \end{subfigure}
    \newline
    \begin{subfigure}{\columnwidth}
        \includegraphics{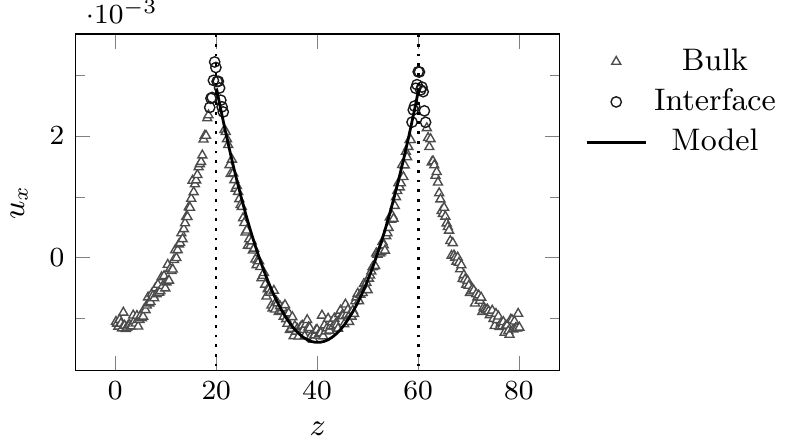}
        \subcaption{$80 \times 80$}
    \end{subfigure}
    \newline
    \begin{subfigure}{\columnwidth}
        \includegraphics{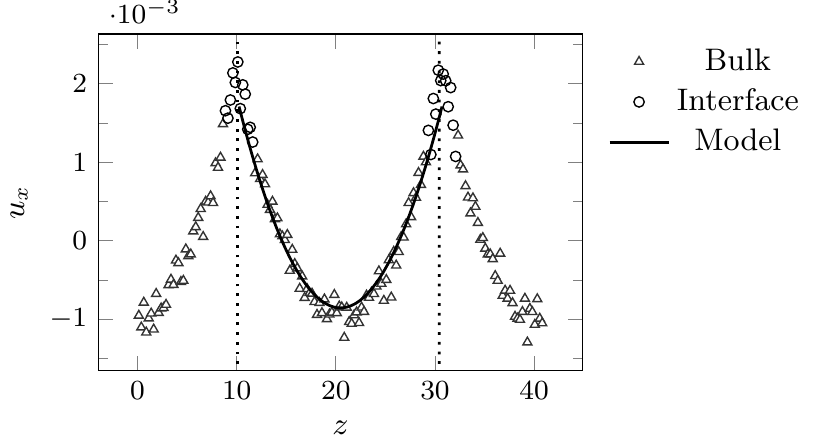}
        \subcaption{$80 \times 40$}
    \end{subfigure}
    \caption{\label{fig:velocity-profiles} Measured velocity profiles $u_x(z)$ along the flow field vortex center positions for different system sizes. Interface centers are marked. Profile calculated with \eqref{eq:velocity-profile} drawn for comparison.}
\end{figure}

\begin{figure}[t]
    \centering
    \begin{subfigure}{\columnwidth}
        \includegraphics{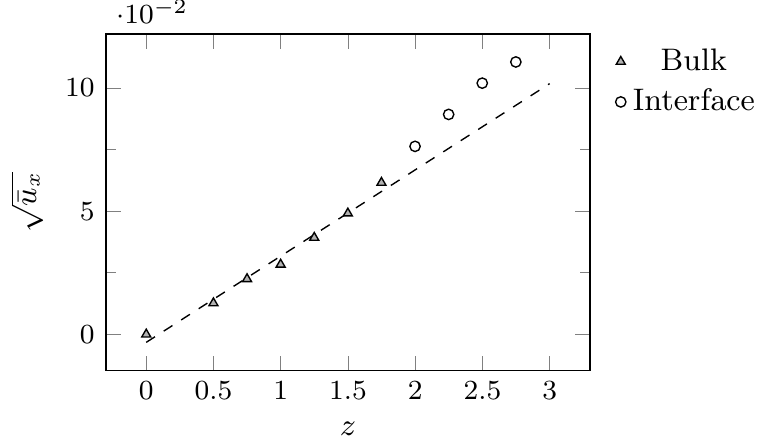}
        \subcaption{$10 \times 10$}
    \end{subfigure}
    \newline
    \begin{subfigure}{\columnwidth}
        \includegraphics{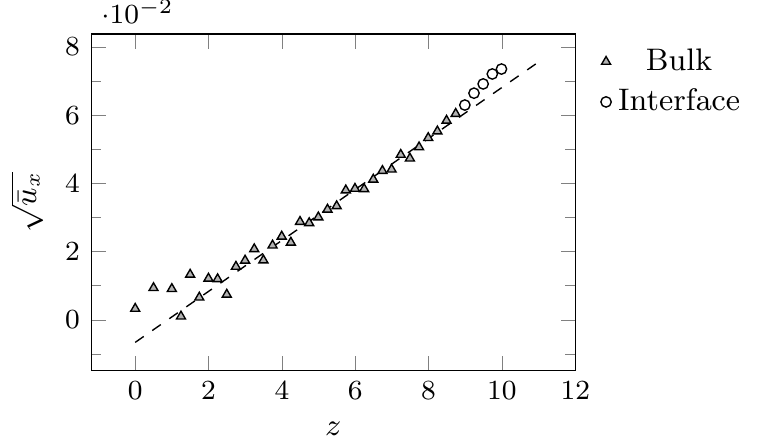}
        \subcaption{$40 \times 40$}
    \end{subfigure}
    \newline
    \begin{subfigure}{\columnwidth}
        \includegraphics{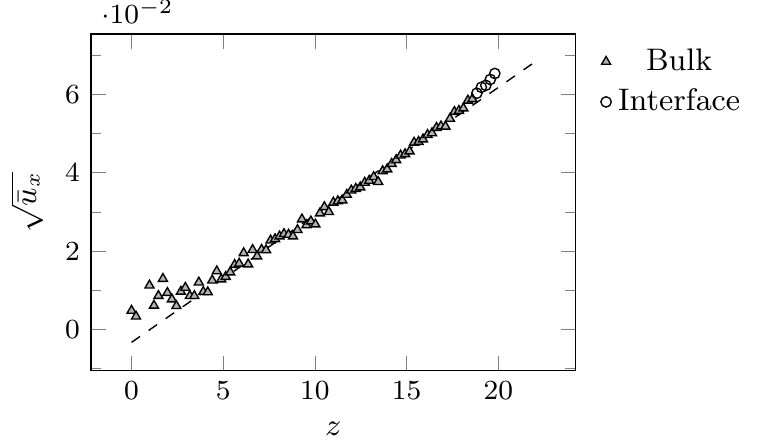}
        \subcaption{$80 \times 80$}
    \end{subfigure}
    \newline
    \begin{subfigure}{\columnwidth}
        \includegraphics{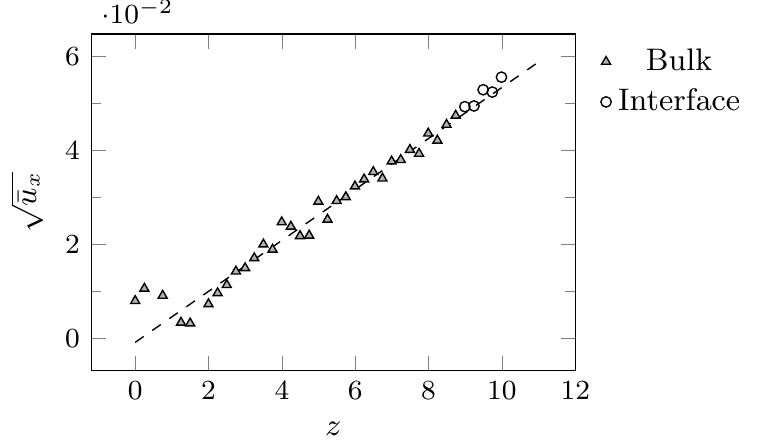}
        \subcaption{$80 \times 40$}
    \end{subfigure}
    \caption{\label{fig:velocity-profiles-sqrt} Profiles of $\sqrt{\bar{u}_x(z)}$ where $\bar{u}_x = u_x(z) - u_0$ shifts the profile to the (approximate) zero velocity. Dashed lines are linear fits in the bulk center.}
\end{figure}

